\documentclass[3p,times,procedia]{elsarticle}
\usepackage{ecrc}
\usepackage[bookmarks=false]{hyperref}
    \hypersetup{colorlinks,
      linkcolor=blue,
      citecolor=blue,
      urlcolor=blue}

\usepackage[table]{xcolor}
\definecolor{maroon}{cmyk}{0,0.87,0.68,0.32}      
\firstpage{1}

\journalname{Procedia Computer Science}

\runauth{M. Spichkova et al.}

\jid{procs}

\usepackage{amssymb}
\usepackage[figuresright]{rotating}

\usepackage{listings} 

\definecolor{codegreen}{rgb}{0,0.6,0}
\definecolor{codegray}{rgb}{0.5,0.5,0.5}
\definecolor{codepurple}{rgb}{0.58,0,0.82}
\definecolor{backcolour}{rgb}{0.95,0.95,0.92}

\lstdefinestyle{mystyle}{
    backgroundcolor=\color{backcolour},   
    commentstyle=\color{codegreen},
    keywordstyle=\color{magenta},
    numberstyle=\tiny\color{codegray},
    stringstyle=\color{blue},
    basicstyle=\ttfamily\footnotesize,
    breakatwhitespace=false,         
    breaklines=true,                 
    captionpos=b,                    
    keepspaces=true,                 
    numbers=left,                    
    numbersep=5pt,                  
    showspaces=false,                
    showstringspaces=false,
    showtabs=false,                  
    tabsize=2
}

\begin{document}
\begin{frontmatter}
\dochead{29th International Conference on Knowledge-Based and Intelligent Information \& Engineering Systems (KES 2025)}%

\title{Advanced approach for Agile/Scrum Process: RetroAI++}

\author[a]{Maria Spichkova*} 
\author[b]{Kevin Iwan} 
\author[b]{Madeleine Zwart} 
\author[a]{Hina Lee}
\author[a]{Yuwon Yoon} 
\author[a]{Xiaohan Qin}  

\address[a]{School of Computing Technologies, RMIT University, Melbourne, Australia}
\address[b]{Shine Solutions, Melbourne, Australia}

\begin{abstract}
In Agile/Scrum software development, sprint planning and retrospective analysis are the key elements of project management.  The aim of our work is to support software developers in these activities. 
In this paper, we present our prototype tool RetroAI++, based on emerging intelligent technologies. In our RetroAI++ prototype, we aim to automate and refine the practical application of Agile/Scrum processes within Sprint Planning and Retrospectives. 
Leveraging AI insights, our prototype aims to automate and refine the many processes involved in the Sprint Planning, Development and Retrospective stages of Agile/Scrum development projects, offering intelligent suggestions for sprint organisation as well as meaningful insights for retrospective reflection. \\ 
\emph{Preprint. Accepted to the 29th International Conference on Knowledge-Based and Intelligent Information \& Engineering Systems (KES 2025). Final version to be published by Elsevier (In Press).} 
\end{abstract}

\begin{keyword}
Software Engineering, Agile, Scrum, Sprint, Sprint planning, AI, Large Language Models, LLMs 
\end{keyword}
\cortext[cor1]{Corresponding author.}
\end{frontmatter}

\email{maria.spichkova@rmit.edu.au}

 \section{Introduction}
 
Agile/Scrum approaches for project management became very popular for software development projects, as they typically are a very good fit for the software development lifecycle (SDL). The popularity of Agile increases with each year, see  \cite{al2020agile,hoda2018rise}. 
The results of a large-scale survey conducted in 2023 by \cite{Agile17}, highlight that 71\% of companies, which participated in the survey, use Agile in their SDL. According to the analysis provided by \cite{schwaber2011scrum}, Scrum is the most popular Agile methodology.  

The key element of Agile is iterative project planning, while in Scrum, these iterations are called \emph{sprints}. 
Sprint planning and analysis of the project progress might be time-consuming. Moreover, many novice software developers struggle with this activity while starting to work in Agile/Scrum settings. For example, a typical challenge we observed is a too high expected velocity. 
Therefore, it would be useful to generate corresponding hints on how well the plan is aligned to Scrum methodology and best practices,  as well as provide automated checks and reports that would support novice developers in following Scrum principles to deliver a software product successfully.
 The aim of our work is to provide software developers with an AI-based support in sprint planning, dependency analysis and corresponding evaluation of the sprint progress.  
  
In our previous work \cite{ENASE25Spichkova}, we presented our preliminary analysis on whether LLMs might be applicable for Scrum retro-meetings, to simplify these meetings and to make them safer psychologically. 
Retro-meetings are a special type of meeting to be conducted at the very end of each development iteration ({sprint}). Their goal is to discuss how the sprint went and to identify what could be done to support continuous improvement within the development team. 
We also introduced the functionality of  \emph{RetroAI++} to support retro-meetings. In this paper, we go further and discuss other fictionalities of our prototype that provide an innovative approach for the Scrum process.

\emph{Contributions:}
In this paper, we discuss challenges that novice software developers encounter while planning the sprints and analysing their progress within the sprints.
We present our prototype tool RetroAI++, based on emerging intelligent technologies, which aims to deal with these challenges. 
In RetroAI++, we aim to automate and refine the practical application of Agile/Scrum processes within Sprint Planning and Retrospectives. Leveraging AI insights, our prototype aims to automate and refine the many processes involved in the Sprint Planning, Development and Retrospective stages of Agile/Scrum development projects, offering intelligent suggestions for sprint organisation as well as meaningful insights for retrospective reflection.

\section{Related work}
\label{sec:related}

In this section, we discuss related works focusing on the 
approaches that support sprint planning and sprint progress analysis, as these two research areas are the focus of RetroAI++ functionality presented in this paper. 
Many of the recent approaches utilise Large Language Models (LLMs).  
A comprehensive literature review on the application of LLMs for Software Engineering (SE) has been presented in  \cite{hou2024large}. The authors identified 395 research articles from January 2017 to January 2024 to categorise different LLMs that have been employed in SE tasks, examined the methods applied in the corresponding studies and approaches, as well as analysed the strategies to optimise and evaluate the performance of LLMs.  Another literature survey on this topic has been presented in \cite{fan2023large}, where the authors especially focused on open research challenges for the application of LLMs to SE.

\textbf{Sprint Planning:}
There were a number of works aiming to support sprint planning. For example, a decision support system for sprint planning in Scrum has been proposed in~\cite{alhazmi2018decision}. 
An optimisation model for multi-sprint planning has been proposed in~\cite{golfarelli2013multi}.
Improvements for the Scrum planning process were proposed in 
\cite{app15010202}. 
In contrast to the above works, our aim is to provide support for software developers within a project planning system. We propose implementing our solution as a prototype tool \emph{RetroAI++}, but it can also be integrated within commercial applications commonly used in the software development industry, e.g., Trello~\cite{trello} or Jira~\cite{jira}. 

Prioritisation of user stories provides background for correct planning: if priorities have been allocated incorrectly, this will impact the overall sprint planning. 
There are also many works on prioritisation of items within a product backlog as well as on requirements prioritisation in general, see for example works in \cite{borhan2022requirements}, \cite{borhan2019requirements}, \cite{jarzkebowicz2020agile}. 
In our study, we analysed the applicability of LLMs for user story prioritisation. 
There are also many research studies on effort estimation, see,  for example, the works by  
\cite{arora2020systematic}, \cite{butt2022software}, \cite{govil2022estimation}.
While the effort estimation also provides a basis for planning, it is out of the scope of our current work. 
In our approach, we assume that effort estimation has already been completed.

A study introduced in \cite{babb2013barriers} presents observations regarding the barriers to learning in the use of Agile methods. The authors conducted two qualitative studies with teams that adapted Agile for software development projects. In our work, we focus on a number of challenges that especially affect novice software developers adapting Scrum.

\textbf{Sprint Progress Analysis and Retrospectives:}
Applicability of speech recognition tools for streamlining the retrospective analysis has been analysed in \cite{gaikwad2019voice}. The authors focused on two tools, Google Home and Amazon Alexa, and investigated potential improvements in the time boxing of a retrospective by using voice-activated commands. This approach might provide a promising extension to RetroAI++ functionality. 

A framework for managing and evaluating changes within the Scrum process has been presented in \cite{hakim2024mped}. The authors didn't focus on providing input for retrospectives, but the framework might be considered for this purpose.

A study presented in \cite{erdougan2018more} analysed how and what kind of historical Scrum project data might be  required for monitoring and  statistical analysis to provide a solid basis for  retrospective meetings, e.g., analysis of the correlation between story
points and actual efforts associated with a product backlog item. 
A recent study \cite{matthies2021experience} aimed to investigate the usage of project data sources in Agile retro meetings, and concluded that a \emph{gather data phase} might be an important part of a retro meeting. 
In our prototype, we suggest going further and providing the data-based input for the retros as part of the RetroAI++ functionality. 

A large-scale and cross-sectional survey \cite{kadenic2023mastering} was conducted to investigate the impact of team maturity and four categories of the Scrum framework (team composition, Scrum values, Scrum roles, and Scrum events) on the perception of being successful at Scrum. This study established a significant correlation between maturity and the perception of being successful at Scrum. 
In our work, we especially aim to support novices, who are especially vulnerable, might be shy to express their thoughts and suggest solutions during the retro meetings. Also, the novices might benefit most from providing additional help and more direct, simple instructions on conducting retros.

\section{Methodology}
\label{sec:methodology}
 
In our RetroAI++ prototype, we aim to automate and refine the practical application of Agile/Scrum processes within Sprint Planning and Retrospectives. Our proposed application, RetroAI++, offers suggestions for sprint organisation as well as insights for retrospective reflection.
This project aims to deliver RetroAI++, which will build upon the existing project to deliver sophisticated Smart Planning features and enhanced Retrospective analysis capabilities.
The prototype combines AI-based planning logic with a more traditional algorithmic foundation in order to enhance the quality of insights produced by the tool. 

Our process has been organised into the following steps, which we discuss in the rest of this section:
\begin{enumerate}
\item 
Analysis of the challenges that novice software developers encounter while working in Agile/Scrum projects.
\item 
Elaboration of algorithmic and AI-based solutions to deal the the above challenges.
\item 
Evaluation of the RetroAI++ on the basis of small-scale studies. 
\end{enumerate}

\subsection{Analysis of the challenges} 
\label{sec:analysis} 

Our observations are based on the following components: (1) Discussions with software engineering practitioners. 
(2) Lessons learned from teaching a course on \emph{Software Engineering Project Management} for Undergraduate and Postgraduate students at RMIT University. The course focuses on Agile/Scrum aspects. The course is large-scale with the number of enrolments above 300, e.g. the number of enrolments in Semester 1 2025 is 525 students.
(3) lessons learned from supervising and mentoring students within their final year software development projects in collaboration with industrial partners.  
We noted that many novice software developers struggle with particular elements of Agile/Scrum. Thus, we observed the following challenges, which we aim to cover in our solutions:
\begin{itemize}
    \item[Ch1:] 
    \emph{Lack of proper ordering of the items in the product backlog.} The items should be sorted by their priorities, while \emph{interdependencies among the items should be taken into account}. Many students tend to allocate lower priority items to the sprints (before higher priority items are completed) or miss the analysis of potential dependencies in the order in which the items should be completed to avoid blockages.  
    \item[Ch2:]
    \emph{Too high expected velocity.}  The expected velocity (the sum of story points of items selected for a sprint) should be close to ideal velocity, but it's acceptable to have it slightly lower. Selecting a too large sprint scope might lead to unnecessary stress towards the end of the sprint. If the team's real velocity is indeed much larger than the current ideal velocity, the team should consider changing the scope of the project or its duration. 
    \item[Ch3:] 
    \emph{Too coarse effort estimation.} 
    If the efforts associated with a single a \emph{Product Backlog Item} (\textbf{PBI}, specified as a user story) are higher than the team's real velocity, this item should be considered as am \emph{epic} and be decomposed in a number of smaller user stories, before allocating any of them to a sprint. 
    This issue is one of the typical contributors to Ch2.  
      \item[Ch4:] 
      \emph{Lack of proper analysis of the team's performance and overall project progress.} The performance and progress have to be analysed at the end of each sprint, but novice developers often struggle with an objective review.  
\end{itemize}
While tools like Trello or Jira allow for facilitating the sprint planning activities remotely, they aim to be a fully flexible application without providing adequate support to novice developers. Neither Trello nor Jira provide any solution to deal with the challenges Ch1, Ch2, and Ch3. Trello also doesn't provide any support to deal with Ch4, while Jira supports sprint analysis in a similar way RetroAI++ does: by generating a sprint burndown chart and a sprint report.  
We propose to incorporate a decision support system into this type of application. To provide a more effective and efficient solution, this system should combine algorithmic and AI-based solutions.

\subsection{Elaboration and Evaluation of algorithmic and AI-based solutions} 

We experimented with algorithmic and AI-based approaches to deal with the challenges discussed in Section~\ref{sec:analysis}. 
To solve challenges Ch1, Ch2, and Ch3, an AI-based approach (implemented in earlier versions of the prototype) demonstrated low accuracy using ChatGPT~3.5. To improve the accuracy and to resource consumption, we simplified the solution to an algorithmic approach. 
To solve the challenge Ch4, the AI-based approach provided reasonable accuracy. We evaluated RetroAI++ on small-scale case studies, where a product backlog includes 20-25 items. 
For all case studies, RetroAI++ provided a reasonable quality sprint summary and feedback. 

To specify a prompt that would allow us to receive a solid and stable response from ChatGPT, we followed the \emph{prompt engineering} methodology. 
The aim of prompt engineering is to optimise LLM input to enhance the output performance, see the work of  \cite{white2023prompt}. 
Our engineering strategy was to elaborate instructions as clearly as possible by adding more explicit constraints to the input. 
In this paper, due to space restrictions, we limit our discussion to the final version of the prompt we specified, see Section~\ref{sec:progress}.


\section{{Implementation of the proposed solution: RetroAI++}}
\label{sec:retroai}

In our prototype, the challenges Ch1, Ch2, and Ch3 are covered by an algorithmic solution (see Section~\ref{sec:SprintPlanning}), which is based on the core parameters like the overall sum of story points in the product backlog, the expected number of sprints, the ideal sprint velocity, etc. This solution uses a classical network diagram~\cite{kruskal1980designing} to specify the interdependencies among the PBIs.
    The information on the dependencies should be added to the product backlog by the software development team.  dependencies. 
For solving Ch4 (see Section~\ref{sec:progress}), we apply the algorithmic solution to create a burn-down chart for the sprint, while an AI-based approach is used to generate a textual report with the analysis of project progress.

The front-end of RetroAI++ has been built using JavaScript and React. For the back-end solution, this project uses Java and DynamoDB tables.  
The prototype runs on AWS and follows the conventions of the Serverless Application Model.  To enhance the performance of Java Lambdas, the project uses native images based on GraalVM built with the assistance of the Quarkus framework in order to significantly reduce the startup time and memory footprint of each serverless function in the application stack. This approach delivered a reduction of 70\% in cold-start time, providing rapid responses from quickly executing functions, which is critical in creating an application that users will want to use. 
In the next section, we would like to discuss in detail the following functionalities of RetroAI++: sprint planning, sprint progress analysis, and project progress visualisation.

\begin{figure}[ht!]
  \centering
  \includegraphics[width=0.48\linewidth]{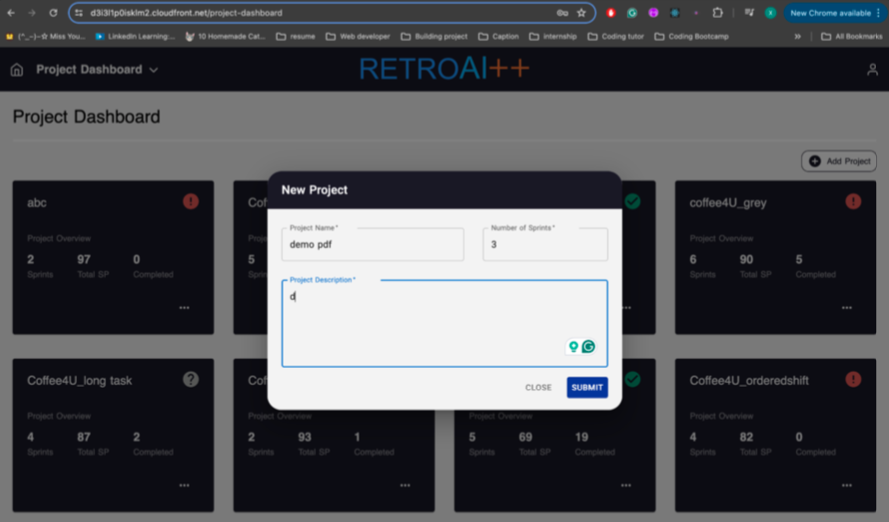} ~~~~~
    \includegraphics[width=0.48\linewidth]{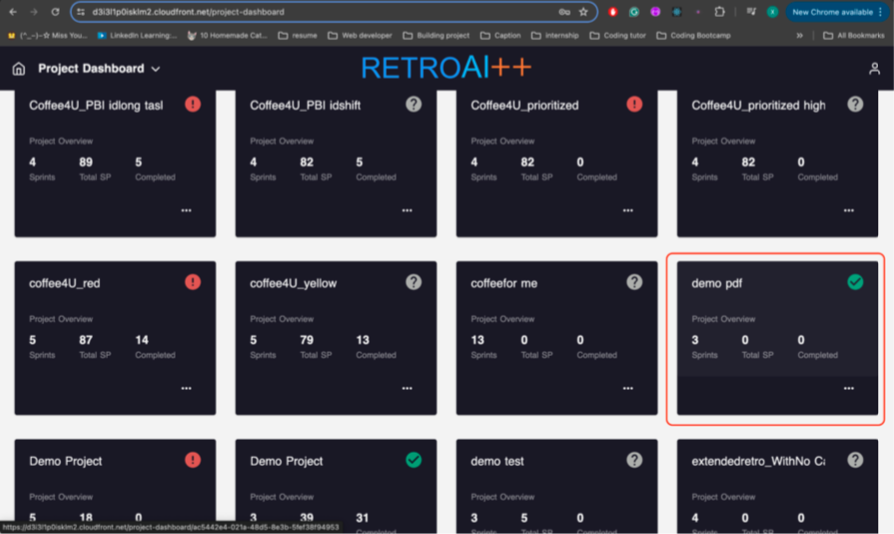} \\
  (a) Form to add a new project ~~~~~~~~~~ ~~~~~~~~~~ 
(b) The new project has been added to the dashboard\\ 
  \caption{RetroAI++: Adding new project to the dashboard.}
  \label{fig:RetroAInew} 
\end{figure}

After the user logs in, the RetroAI++ dashboard provides an overview of all the user's projects. 
Figure~\ref{fig:RetroAInew} presents the project creation functionality and the dashboard after the new project has been created.   
RetroAI++ provides tool support for sprint planning and retrospective analysis, including facilitation of retrospective meetings (retros), but in this paper, we focus on its functionality dedicated to the sprint planning and sprint progress analysis.

 \subsection{RetroAI++: Sprint planning}
 \label{sec:SprintPlanning}

\begin{figure}[ht!]
  \centering
  \includegraphics[width=0.48\linewidth]{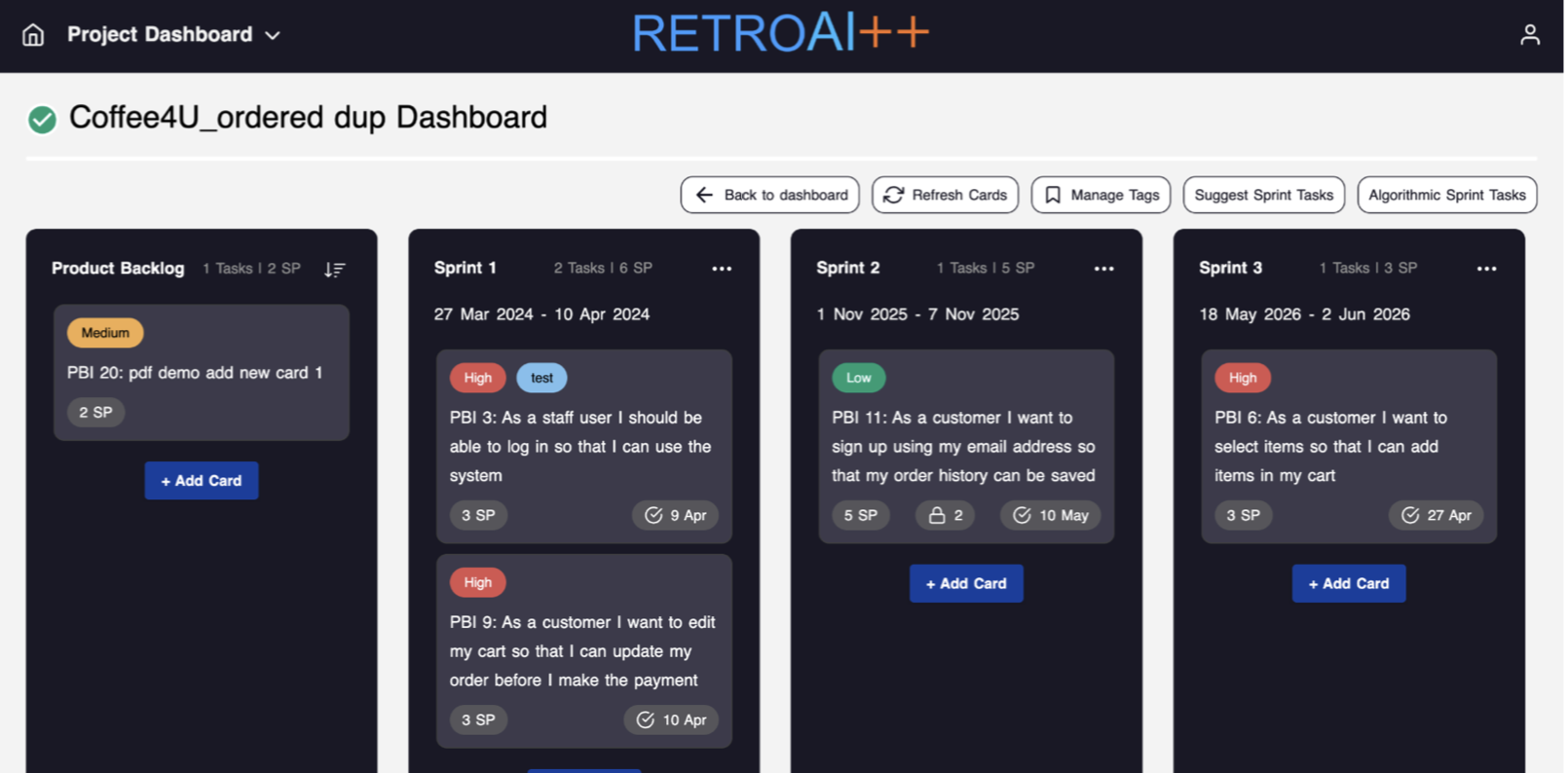} 
  ~~~~~
    \includegraphics[width=0.48\linewidth]{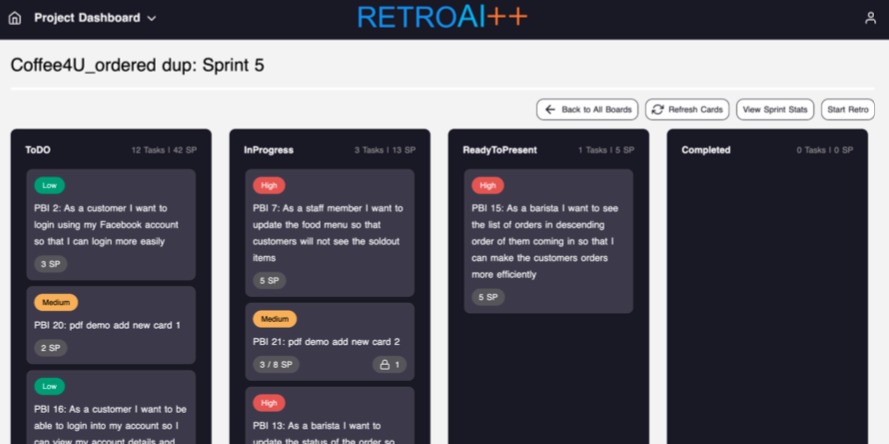}\\
    (a) Project overview page
~~~~~~~ ~~~~~~~ ~~~~~~~ 
  ~~~~~~~ ~~~~~~~ ~~~~~~~
    (b) Sprint board
  \caption{RetroAI++: Project overview page and Sprint board.}
  \label{fig:RetroAIplanning} 
\end{figure}

Figure~\ref{fig:RetroAIplanning} demonstrates a page presenting an overview of a project and a sprint board for a project. 
 The Sprint Planning Algorithm is designed to enhance the efficiency and accuracy of sprint planning within an Agile/Scrum framework. It leverages priority-based sorting and dependency resolution to ensure optimal task assignment across sprints.

 Our solution considers the following constraints, which allow us to deal with the challenges Ch1, Ch2, and Ch3.

  (C1)   Lower priority cards must not precede higher priority cards, i.e.
     \begin{equation}
      for~all~i,j  \in PB:~ priority_i > priority_j \to preceed(i, j)
 \end{equation}
 where $priority_i$ denotes priority of the PBI $i$. 
 Applying this constraint allows RetroAI++ to eliminate the challenge Ch1 (\emph{Lack of proper ordering of the items in the product backlog}). \\

  (C2)  
    The expected velocity (the sum of story points of PBIs selected for a sprint) must not exceed the ideal velocity, i.e., the expected velocity must be less than or equal to the ideal velocity.
        \begin{equation}
      for~all~k  \in \textit{Sprints}:  
      v_{expected}(k)  \le v_{ideal}
    \end{equation}
    \[
    v_{expected}(k) = \sum_{i \in \textit{Selected}(k)} e_i\]
where \textit{Selected} denotes the set of items selected for a sprint, 
    $\textit{Selected}(k) \subset PB$.
    \\ 
Applying this constraint allows RetroAI++ to eliminate the challenge Ch2 (\emph{Too high expected velocity}).\\

(C3)  
 Efforts $e_i$ (specified in story points) of a single PBI $i$ must not exceed the ideal sprint velocity $v_{ideal}$ expected for this project, i.e.
 \begin{equation}
      for~all~j  \in PB:~ e_i \le v_{ideal}
 \end{equation}

 \[v_{ideal} = (\sum_{j \in PB} e_j) / N_s
 \]
where $PB$ denotes the set of all items in the product backlog, and $N_s$ denotes the expected number of sprints within the project.   
    Thus, if a PBI is not a simple user story but an epic, the user will get a corresponding warning/recommendation to decompose an item or label it as an epic. 
Applying this constraint allows RetroAI++ to eliminate the challenge Ch3 (\emph{Too coarse effort estimation}).\\

 (C4)  
    Sprints must have valid start and end dates. End date cannot precede the start date, i.e. 
    \begin{equation}
      for~all~k  \in \textit{Sprints}:  startDate(k) < enddate(k)
    \end{equation}
    where \textit{Sprints} denotes the set of all sprints in the project. 
While this constraint doesn't eliminate any core challenges that we observed, it helps avoid confusion in the specification of the sprint dates.

 \subsection{RetroAI++: User story dependency and priority validation}

The proposed algorithm includes robust mechanisms for validating dependencies and priorities of tasks to ensure the integrity of the sprint planning process. This validation prevents circular dependencies and ensures that the priority of tasks adheres to predefined rules. Specifically, a high-priority task cannot be blocked by a lower-priority task. 
 
\begin{figure}[ht!]
  \centering
  \includegraphics[width=0.48\linewidth]{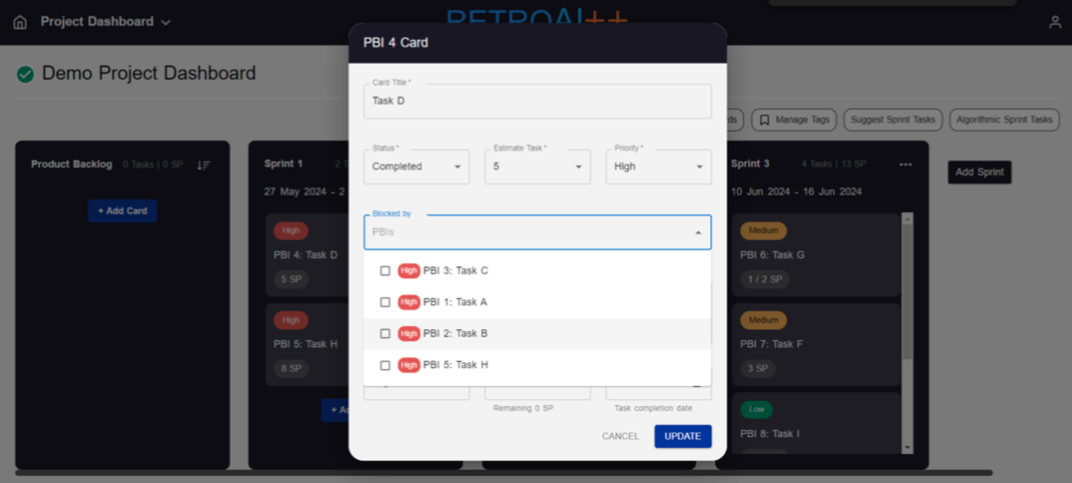} ~~~~~ 
  \includegraphics[width=0.48\linewidth]{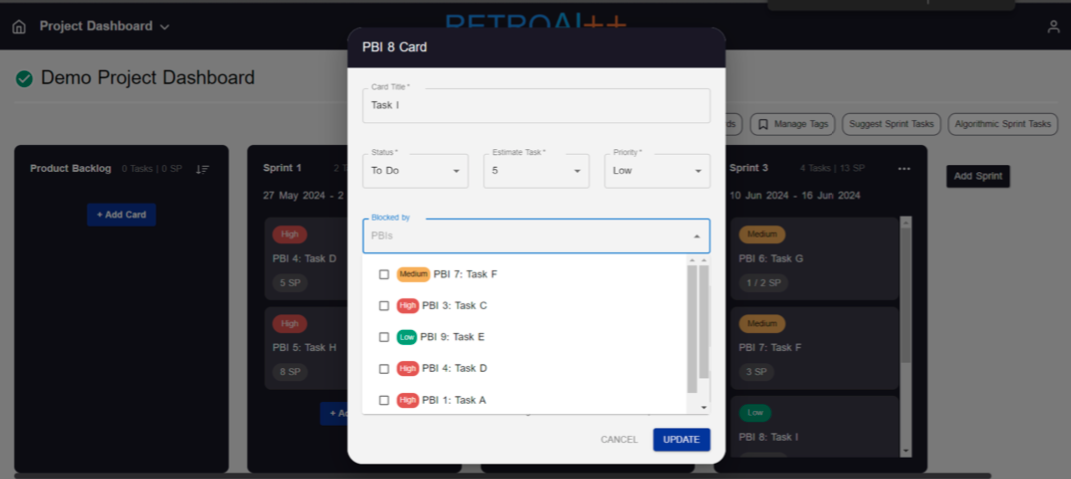}\\
  (a) dependency specification for a \emph{High} priority PBI  ~~~~~ 
   (b) dependency specification for a \emph{Low} priority PBI  
  \caption{RetroAI++: Specification of dependencies among Product Backlog Items (PBIs).}
  \label{fig:RetroAIdep1} 
\end{figure}

 Let $dependency(i,j)$ denotes that the item $i$ depends on the item $j$, and $cannotBlock(j,i)$ denotes that the item $j$ cannot block the item $i$, i.e. $cannotBlock(j,i) = \nexists~ dependency(i,j)$, then 
 \begin{equation}
     for~all~i,j  \in PB:~ priority_i > priority_j \to cannotBlock(j,i)
 \end{equation} 
 Figure~\ref{fig:RetroAIdep1} presents two examples of dependency specification in the project: 
 \begin{itemize}
  \item[(a)] 
    The selected PBI has \emph{High} priority, therefore, it can depend only on items with  \emph{High} or \emph{Critical} priority.
   \item[(b)] 
     The selected PBI has \emph{Low} priority, therefore, it can depend on items with \emph{Critical}, \emph{High}, \emph{Medium} or \emph{Low} priority, i.e., on any item in the project except itself.
 \end{itemize}

 Correspondingly, a PBI cannot change its priority to lower than the items depending on it (i.e., the items blocked by this PBI):
 \begin{equation}
          for~all~i,j  \in PB: dependency(i, j) \to  priority_i \le priority_j  
 \end{equation}
 
 Similarly, a PBI cannot change its priority to higher than the items it depends on. This ensures that items with higher priority are not dependent on items with lower priority, maintaining the proper workflow and task management. Moreover, mutual/circular dependencies are forbidden: 
  \begin{equation}
          for~all~i,j  \in PB: dependency(i, j) \to  \neg dependency(j, i)
 \end{equation}
 
\begin{figure}[ht!]
  \centering
  \includegraphics[width=0.48\linewidth]{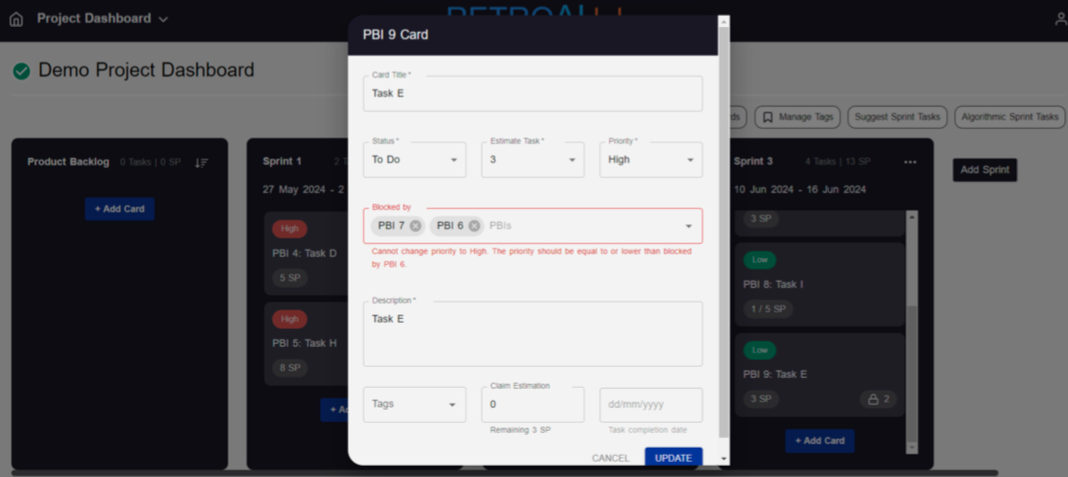}   ~~~~ \includegraphics[width=0.48\linewidth]{RetroAIdep2.png}
  \caption{RetroAI++: Application of constraints on updates for PBI priority and the corresponding hints to follow the constraints.}
  \label{fig:RetroAIdep3} 
\end{figure}

 If a user tries to update the priority violating the above constraints, the corresponding error message will be displayed, see Figure~\ref{fig:RetroAIdep3}. 
 \\
 ~
 
 \subsection{RetroAI++: Sprint progress analysis} 
 \label{sec:progress}

\begin{figure}[ht!]
  \centering
  \includegraphics[width=0.47\linewidth]{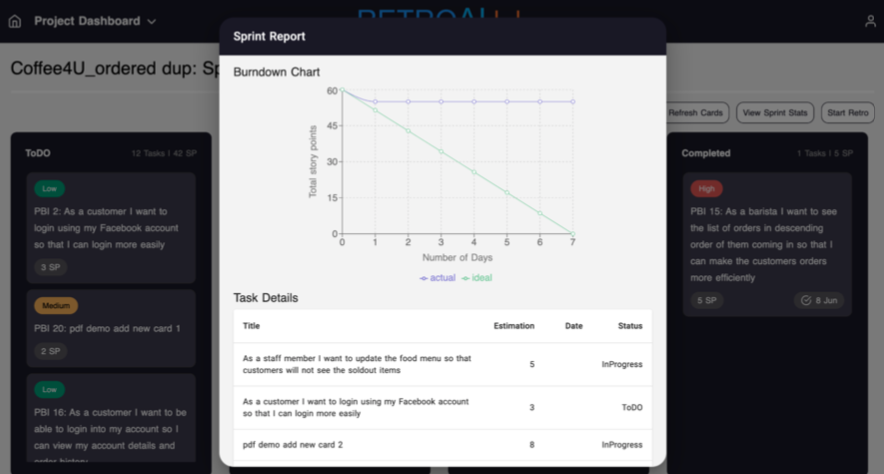}
  ~~~~~~
  \includegraphics[width=0.45\linewidth]{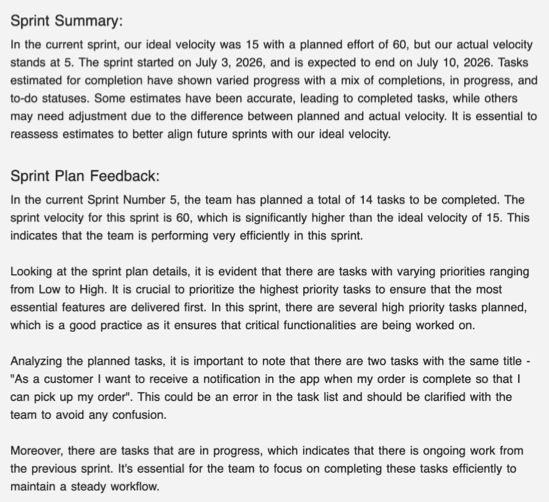}
  \caption{RetroAI++: Sprint burn-down chart and AI-generated Sprint summary.}
  \label{fig:RetroAIburndown} 
\end{figure}

In this subsection, we present our solution to deal with the Challenge Ch4 (\emph{Lack of proper analysis of the team’s performance and overall project progress}).    
  Figure~\ref{fig:RetroAIburndown}  demonstrates a screenshot of the RetroAI++  page, 
 where a burn-down chart for a sprint has been presented,  
   as well as an AI-generated textual report of the project progress within the sprint. In the latest version of our prototype, we used the capacity of OpenAI ChatGPT 4.0 to implement this functionality. 
 
Using the \emph{prompt engineering} methodology, we elaborated a prompt that allows for obtaining the following:
 \begin{itemize}
  \item[(a)] 
    A brief \emph{Sprint summary} that provides an overview of a sprint, covering the core aspects of Scrum project planning and management. We limited the length of the summary to 150 words to provide an easy-to-read overview that can be useful for starting the discussion in the \emph{Sprint Review} meeting.
  \item[(b)]
    A \emph{Sprint Plan Feedback} that provides an analysis of the achieved results with respect to the sprint backlog initially created for the sprint; this analysis can be a good starting point for a discussion within the \emph{Sprint Retrospective} meeting.
 \end{itemize}
The code of our ChatGPT query is presented as a listing in Figure~\ref{fig:code}. We shared with ChatGPT the list of tasks selected for the sprint (along with their statuses, priorities and story point estimates), the ideal velocity $v_{ideal}$, the expected velocity $v_{expected}$ (planned effort), and the actual (real) velocity within the sprint. The real velocity is the sum of the story points of all PBIs completed within the sprint. 

    \begin{figure}[ht!]
  \centering
  \begin{minipage}{0.95\linewidth}
\begin{lstlisting}[language=Java]
        String generateReportQuery = "Write a 150 word long summary about a sprint with the following details, including planned vs actual statistics."
                +
                "Mention where estimates were accurate and where they should be adjusted. Give an introduction and conclusion. Do not give a task by task break down.";

        String calculatedEffortQuery = "ideal velocity: " + String.format("%.0f", velocity) + "\\n" +
                "planned effort: " + estimatedEffort + "\\n" +
                "actual velocity: " + actualVelocity + "\\n" +
                board.toStringForReport();

        String tasksQuery = String.join(", ", tasks.stream()
                .map(Task::toStringForReport)
                .collect(Collectors.toList()));

        String finalQuery = generateReportQuery + calculatedEffortQuery + "\\n Tasks: \\n" + tasksQuery;

        String generatedReport = callChatGPTAPI(gptApiUrl, gptApiKey, finalQuery, "1"); 
\end{lstlisting}
\end{minipage}
  \caption{Code sample of RetroAI++: Functionality to apply OpenAI ChatGPT capacity to generate a Sprint summary.}
  \label{fig:code} 
\end{figure} 

 \subsection{RetroAI++: Project performance status}
An important element of the sprint progress discussion is the analysis of the team's performance with respect to expected performance for the current project iteration (sprint) and for the project in general. In Scrum, this is typically done by comparison of the real (actual) team's velocity vs. ideal and acceptable velocities. In RetroAI++, we propose to visualise the project performance status using traffic light annotations that should be visible both on the dashboard and on the project board. 

\begin{figure*}[ht!]
  \centering
  ~\\
  ~\\
  \includegraphics[width=0.48\linewidth]{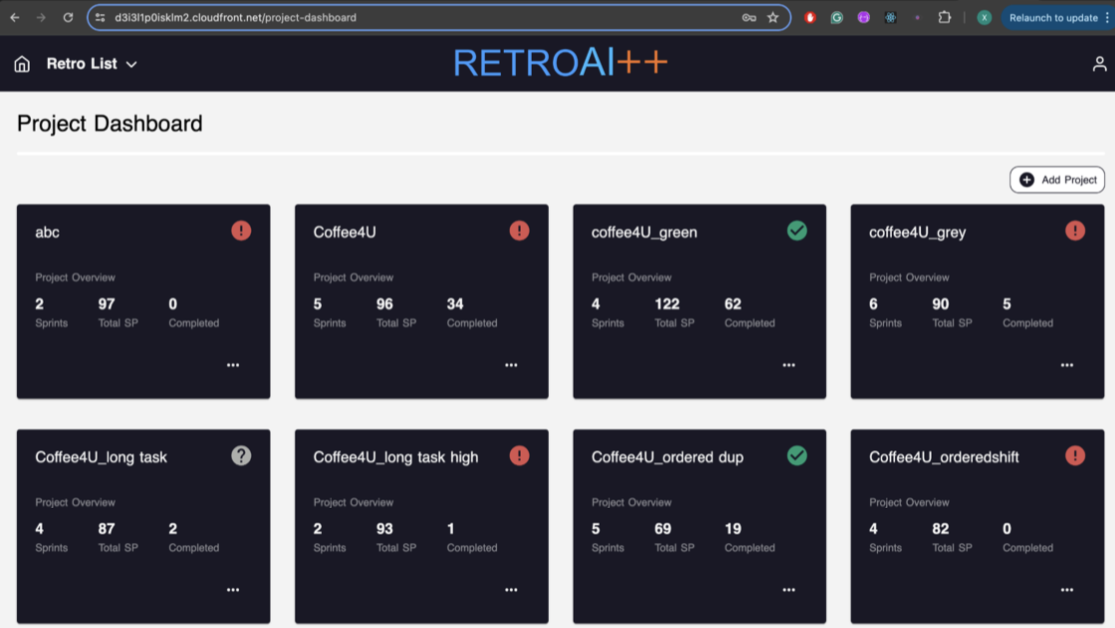}~~~~~ \includegraphics[width=0.48\linewidth]{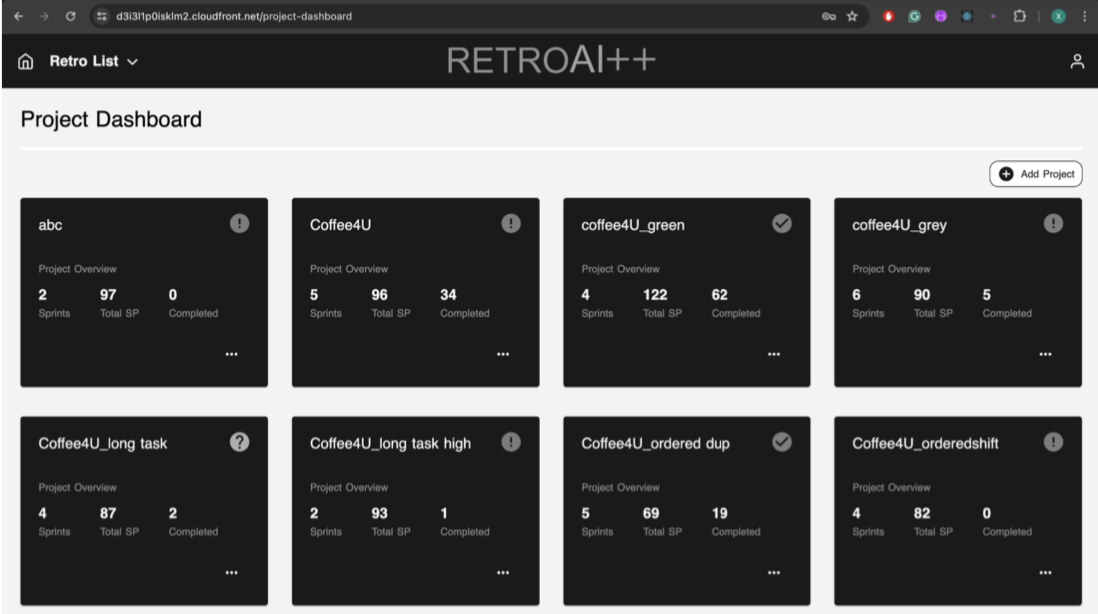}  
\\
  (a) Trichromacy/ Normal \hspace{30mm} (b) Monochromacy/ Achromatopsia\\
  \vspace{2mm}
   \includegraphics[width=0.48\linewidth]{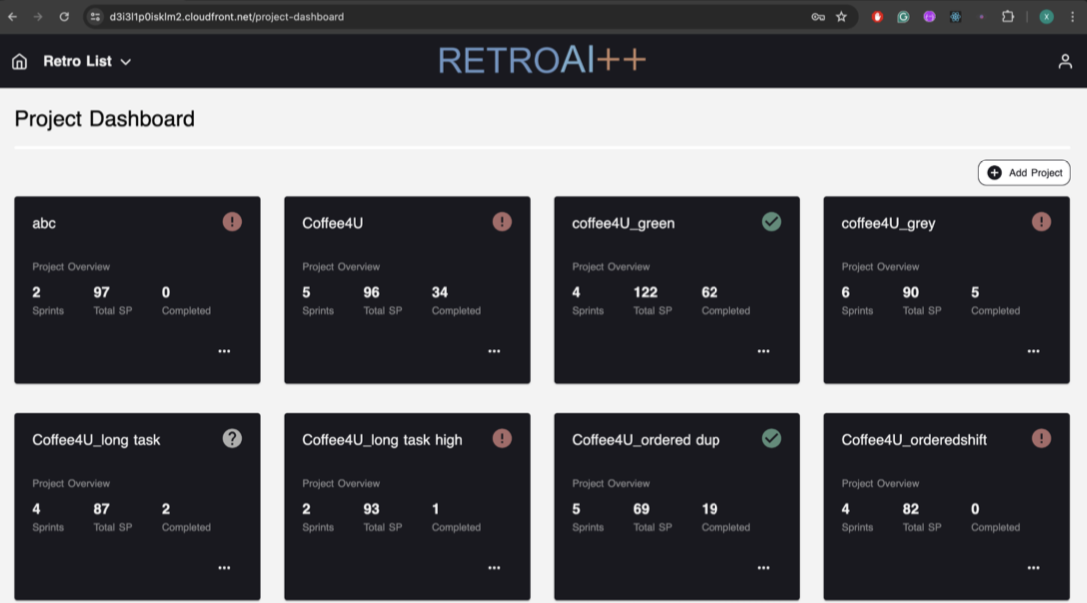} ~~~~~  \includegraphics[width=0.48\linewidth]{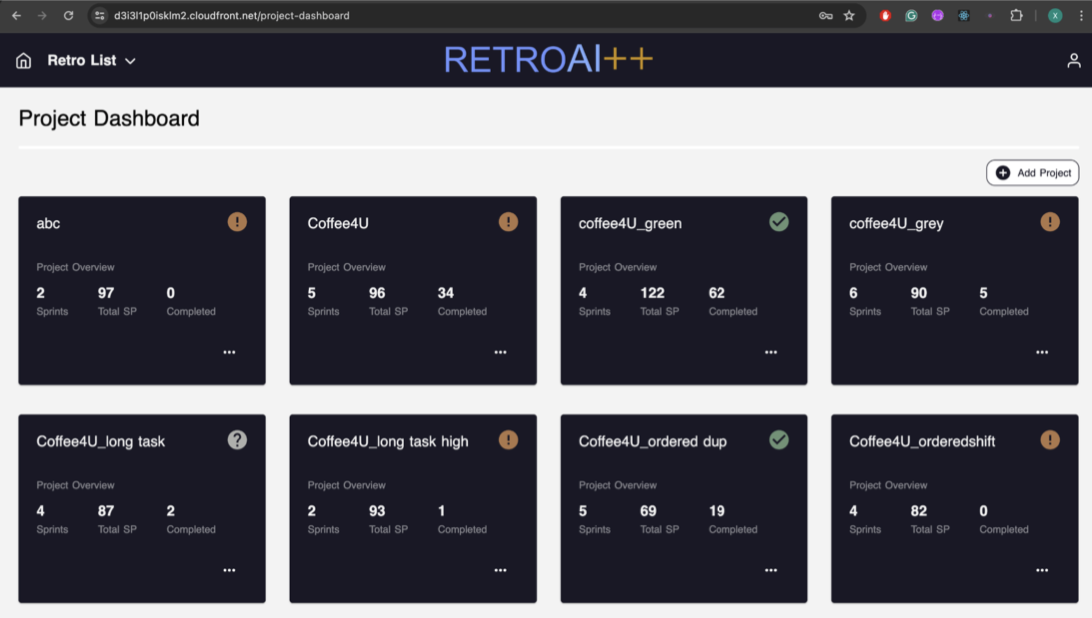}
   \\
   (c) Blue Cone Monochromacy/ Achromatomaly \hspace{30mm} (d) Green-Blind/ Deuteranopia
   
   \caption{RetroAI++: Project dashboard presented for different colour vision categories (dark version of the UI), where the key indicators of the projects' statuses are highlighted according to the colour schema that is visually distinguishable for the corresponding colour vision categories}
  \label{fig:RetroAI-dashboard-overview} 
\end{figure*}

To increase the usability of a software interface, it's important to provide solutions for colour blind users~\cite{jefferson2006accommodating}. 
The core usability metrics are specified by Nielsen~\cite{nielsen1996usability} as follows: 
(1) success rate, i.e., whether users can perform the task at all, (2) the time a task requires, (3) the error rate, and
(4) users' subjective satisfaction. 
Our prototype takes into account diversity in colour vision categories~\cite{colourblindawareness},   such as 
Trichromacy/ Normal, 
 Monochromacy/ Achromatopsia,  
Green-Weak/ Deuteranomaly,  
Green-Blind/ Deuteranopia,  
Red-Weak/ Protanomaly,  
Red-Blind/ Protanopia,  
Blue Cone Monochromacy/ Achromatomaly, 
Blue-Weak/ Tritanomaly, and 
Blue-Blind/ Tritanopia. This doesn't impact the success rate of using our tool and rate of critical errors, but it impacts 
the time the developer will require to analyse the dashboard and, respectively, their subjective satisfaction with the tool. 
Figure~\ref{fig:RetroAI-dashboard-overview} illustrates this on the example of RetroAI++ project dashboard page, where the overall project performance in terms of team's velocity. The indicators of the project performance are not only presented within the required colour schema but also have other visual hints to decrease the cognitive load:
``$\checkmark$'' to denote projects being on track,  ``!'' to denote projects having significant delays (i.e., being at risk), ``?'' to denote projects having some delays and requiring additional attention to avoid being at risk. 

\section{Conclusions}
\label{sec:summary}	

In this paper, we presented our ongoing research on streamlining Agile/Scrum processes with the support of AI approaches.   
We presented our prototype tool, RetroAI++, whose aim is to automate and simplify Agile/Scrum processes for software development projects. We especially focused on AI-enhanced RetroAI++ functionality for sprint planning and monitoring of project progress. 

  This research project has been conducted under the initiative 
\emph{Research embedded in teaching}, see, for example,  \cite{chugh2019automated,george2020usage,spichkova2019industry,spichkova2020vm2,spichkova2017autonomous,spichkova2020ICSoft}.  
This initiative has been introduced in RMIT University (Melbourne, Australia)   to encourage students' curiosity for Software Engineering and Computer Science research, focusing on emerging technologies that might improve industrial processes and/or practices.

\emph{Limitations:} The RetroAI++ prototype has a number of limitations, which we aim to cover in our future work. 
(1) The current algorithmic solution covers only a limited set of constraints, focusing on the core challenges we observed among novice developers. 
(2) The sprint progress is analysed based only on the information presented in the Sprint board, while it might also be useful to include information from other sources, e.g., issue trackers. 
(3) RetroAI++ currently creates only \emph{sprint} burndown charts (burndown charts covering a single sprint). This is useful to analyse the progress within a sprint, but it doesn't help in the analysis of the overall progress within the project.  It would be useful to have a \emph{project} burndown chart as well. 
(4) The prototype has been evaluated in small-scale studies. An additional evaluation based on a larger study would be beneficial to have. 

 
\emph{Future work:}
As our future work, we plan to conduct experiments on expanding the algorithmic solution.   
Another potential direction of our future work is to expand the sprint summary (as the starting point of a retro-discussion) by including more detailed information from issue trackers. We also plan to implement a feature to generate a project burndown chart. 
Another promising direction for future work would be the integration of RetroAI++ with speech recognition tools for streamlining the retrospective analysis.

\section*{\uppercase{Acknowledgements}}

We would like to thank Shine Solutions for sponsoring this project under the research grant PRJ00002505, and especially Branko Minic and Adrian Zielonka for sharing their industry-based expertise and advice. We also would like to thank students who contributed to the creation of earlier versions of the RetroAI tool: 
Weimin Su, Ahilya  Sinha,  
Hibbaan  Nawaz, 
Kartik Kumar,  
Muskan Aggarwal,  
Justin John, Shalvi Tembe, Niyati Gulumkar, Vincent Tso, and Nguyen Duc Minh Tam. 

\bibliographystyle{elsarticle-harv}

\end{document}